\documentstyle[twoside,fleqn,espcrc2]{article}
\newread\epsffilein    
\newif\ifepsffileok    
\newif\ifepsfbbfound   
\newif\ifepsfverbose   
\newdimen\epsfxsize    
\newdimen\epsfysize    
\newdimen\epsftsize    
\newdimen\epsfrsize    
\newdimen\epsftmp      
\newdimen\pspoints     
\def\epsfbox#1{\global\def\epsfllx{72}\global\def\epsflly{72}%
   \global\def\epsfurx{540}\global\def\epsfury{720}%
   \def\lbracket{[}\def\testit{#1}\ifx\testit\lbracket
   \let\next=\epsfgetlitbb\else\let\next=\epsfnormal\fi\next{#1}}%
\def\epsfgetlitbb#1#2 #3 #4 #5]#6{\epsfgrab #2 #3 #4 #5 .\\%
   \epsfsetgraph{#6}}%
\def\epsfnormal#1{\epsfgetbb{#1}\epsfsetgraph{#1}}%
\def\epsfgetbb#1{%
\openin\epsffilein=#1
\ifeof\epsffilein\errmessage{I couldn't open #1, will ignore it}\else
   {\epsffileoktrue \chardef\other=12
    \def\do##1{\catcode`##1=\other}\dospecials \catcode`\ =10
    \loop
       \read\epsffilein to \epsffileline
       \ifeof\epsffilein\epsffileokfalse\else
          \expandafter\epsfaux\epsffileline:. \\%
       \fi
   \ifepsffileok\repeat
   \ifepsfbbfound\else
    \ifepsfverbose\message{No bounding box comment in #1; using defaults}\fi\fi
   }\closein\epsffilein\fi}%
\def\epsfclipstring{}
\def\epsfsetgraph#1{%
   \epsfrsize=\epsfury\pspoints
   \advance\epsfrsize by-\epsflly\pspoints
   \epsftsize=\epsfurx\pspoints
   \advance\epsftsize by-\epsfllx\pspoints
   \epsfxsize\epsfsize\epsftsize\epsfrsize
   \ifnum\epsfxsize=0 \ifnum\epsfysize=0
      \epsfxsize=\epsftsize \epsfysize=\epsfrsize
      \epsfrsize=0pt
     \else\epsftmp=\epsftsize \divide\epsftmp\epsfrsize
       \epsfxsize=\epsfysize \multiply\epsfxsize\epsftmp
       \multiply\epsftmp\epsfrsize \advance\epsftsize-\epsftmp
       \epsftmp=\epsfysize
       \loop \advance\epsftsize\epsftsize \divide\epsftmp 2
       \ifnum\epsftmp>0
          \ifnum\epsftsize<\epsfrsize\else
             \advance\epsftsize-\epsfrsize \advance\epsfxsize\epsftmp \fi
       \repeat
       \epsfrsize=0pt
     \fi
   \else \ifnum\epsfysize=0
     \epsftmp=\epsfrsize \divide\epsftmp\epsftsize
     \epsfysize=\epsfxsize \multiply\epsfysize\epsftmp   
     \multiply\epsftmp\epsftsize \advance\epsfrsize-\epsftmp
     \epsftmp=\epsfxsize
     \loop \advance\epsfrsize\epsfrsize \divide\epsftmp 2
     \ifnum\epsftmp>0
        \ifnum\epsfrsize<\epsftsize\else
           \advance\epsfrsize-\epsftsize \advance\epsfysize\epsftmp \fi
     \repeat
     \epsfrsize=0pt
    \else
     \epsfrsize=\epsfysize
    \fi
   \fi
   \ifepsfverbose\message{#1: width=\the\epsfxsize, height=\the\epsfysize}\fi
   \epsftmp=10\epsfxsize \divide\epsftmp\pspoints
   \vbox to\epsfysize{\vfil\hbox to\epsfxsize{%
      \ifnum\epsfrsize=0\relax
        \includegraphics{#1}%
      \else
        \epsfrsize=10\epsfysize \divide\epsfrsize\pspoints
        \includegraphics{#1}%
      \fi
      \hfil}}%
\global\epsfxsize=0pt\global\epsfysize=0pt}%
{\catcode`\%=12 \global\let\epsfpercent=
\long\def\epsfaux#1#2:#3\\{\ifx#1\epsfpercent
   \def\testit{#2}\ifx\testit\epsfbblit
      \epsfgrab #3 . . . \\%
      \epsffileokfalse
      \global\epsfbbfoundtrue
   \fi\else\ifx#1\par\else\epsffileokfalse\fi\fi}%
\def\epsfempty{}%
\def\epsfgrab #1 #2 #3 #4 #5\\{%
\global\def\epsfllx{#1}\ifx\epsfllx\epsfempty
      \epsfgrab #2 #3 #4 #5 .\\\else
   \global\def\epsflly{#2}%
   \global\def\epsfurx{#3}\global\def\epsfury{#4}\fi}%
\def\epsfsize#1#2{\epsfxsize}
\let\epsffile=\epsfbox

\newcommand{\AmS}{{\protect\the\textfont2
  A\kern-.1667em\lower.5ex\hbox{M}\kern-.125emS}}

\hyphenation{author another created financial paper re-commend-ed}

\title{Confining Classical Configurations}

\author{A. Gonz\'alez-Arroyo and A. Montero \address{Dpto. de F\'{\i}sica Te\'orica C-XI, 
        Univ. Aut\'onoma de Madrid, \\ 
        Cantoblanco, Madrid 28049, SPAIN}%
        \thanks{Work financed by CICYT  grant AEN93-0693}
   }
       
\begin{document}

\begin{abstract} 
We construct a family of smooth, almost self-dual, non-thermalized  SU(2) gauge field configurations, and measure the average of 
the fundamental, adjoint and spin  $\frac{3}{2}$ representation
Wilson loops  on them. We get area law in all three cases. We also study thermalised configurations at 
$\beta= 2.325$ after cooling. The ratio of string tension in the spin j representation over that in the fundamental, stays
constant with cooling. 

\end{abstract}

\maketitle

\section{Introduction}
Can classical configurations produce Confinement? What we mean with this question is whether  one can produce an 
statistical distribution of smooth configurations such that the expectation value of Wilson loops  on them shows 
an area law behaviour. The question makes sense since there are different authors which defend  opposite 
answers to this question. The question is preliminary to a much more important one: Can one understand 
Confinement in QCD in terms of classical configurations? The general idea was advocated by Polyakov 
in the 70's. Since then, many authors have defended this point of view, although  proposing different alternatives  for what
the relevant physical configurations can be. Our group \cite{invest} proposed that the configurations are smooth, 
quasi-self-dual, 
and dense enough so that instantons loose their individuallity. Furthermore, we suggested that in this situation 
fractional topological charge objects begin to be identifiable structures and form a liquid, which explains many properties 
of the Yang-Mills vacuum, and in particular Confinement.  
Although at present we cannot provide conclusive evidence in favour of our  model, we will be able to give results 
which help in solving the first two questions. In addition, following a suggestion by J. Ambjorn, we have studied the string tension
in the adjoint and spin=3/2 representation, a point which has been advocated by Greensite \cite{greensite} to be an important test on
different models of Confinement. We will dedicate the next  sections to present these results.
See also Ref. \cite{doclass}.

\section{Can classical configurations produce Confinement?}

In order to answer this question we  generated a family of SU(2) gauge 
configurations and measured the string tension on them. To generate them, we proceed as 
follows. We start with a  configuration on a $16^3 \times 8$
lattice and periodic boundary conditions, built  by gluing a $4^4$ building block 
to itself. This building block is the minimum action configuration on a  $4^4$ lattice
with twisted boundary conditions $\vec{m} = \vec{k} = (1,1,1)$. It is self-dual 
and has topological charge $Q = 1/2$. The final configuration is a sort of crystal 
of these lumps. To {\em melt} the crystal we applied a series of Monte Carlo 
steps at a large value of $\beta$. We take the minimum  number of Monte Carlo steps necessary
to produce  a fairly uniform distribution of lumps. This number grows with $\beta$, being 50 at 3.2 and 75 at 3.8. 
In any case the resulting configuration is far from being thermalised. Finally, we apply the cooling method 
of Ref. \cite{overimprove} with $\epsilon = -0.3$.  The whole  procedure is repeated several times 
to obtain a family of configurations.  We have studied several values  of  $\beta$ in the range [3, 5] with identical 
conclusions. Here we will present the results for  $\beta = 3.8$, for which we have a total 
of 150 configurations. We have studied the dependence of the results on the number of applied cooling
 steps. We only find a significant difference during the first few cooling steps. Beyond  20 cooling
 steps the results vary very little. 

\begin{figure}
\begin{center}
\leavevmode
\epsfverbosetrue
\epsfxsize=200pt
\epsfysize=115pt
\epsffile{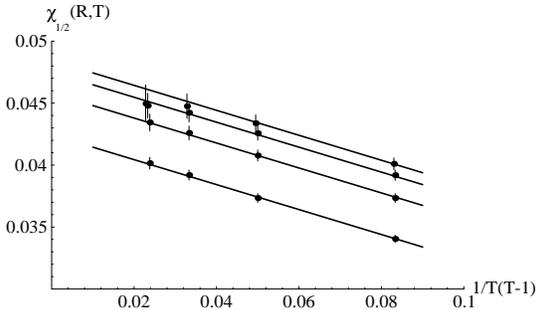}
\caption{ The values of $\chi_{\frac{1}{2}}(R,T)$ at 20 cooling steps for R=7,6,5,4 (top to bottom) as a function of $\frac{1}{T(T-1)}$
 compared with the straight  line prediction of Eq.~1.}
\end{center}
\end{figure}

Once we got this sample of configurations, we computed the average value, over positions and configurations,
of square $R \times T$ Wilson loops, and obtained the Creutz ratios from them. For a smaller subsample
of configurations we have also measured loops and Creutz  ratios for the adjoint  (110 configurations)
and the spin$=3/2$ representation (50 configurations). We have fitted our data on Creutz ratios to the formula: 
\begin{equation}
\chi_j(R,T) =  \sigma_j  -  \gamma_j ( \frac{1}{R(R-1)} + \frac{1}{T(T-1)}) \  ,
\end{equation}  
where $\sigma_j$ is the string tension in the spin $j$ representation,
 and the term proportional to $\gamma_r$ follows from a scale-invariant contribution to
Wilson loops (perturbation theory or string fluctuations). The main conclusions from our results are the following:

1.  The fits, done over a range of $R$ and $T$ values going from 4 to 7,  are excellent, giving  $\chi^2$ values
per degree of freedom which are much smaller than 1 (the values for different sizes are correlated). In Fig. 1,
we show, for 20 cooling steps, the data of $\chi_{\frac{1}{2}}(R,T)$ for R,T =4, 5, 6 and 7 and plotted as a 
function of 1/(T(T-1)), together 
with the curves of our 2 parameter fit.
 
2. There is a non-zero value of the string tension in all three representations in the  range of distances probed.
 For 10 cooling steps the 3  
string tensions are in the ratio 1:1.83:2.7  and
drops to 1:1.53:1.93 for 100 cooling steps.

3. The value of the fundamental string tension decreases with cooling but  seems to approach a constant. For example,
 $\sigma_{\frac{1}{2}}$ is equal to 0.0554, 0.0508, 0.0461 and 0.0453 for 10, 20, 50 and 100 coolings respectively.

In conclusion, we have shown how our configurations provide an affirmative answer to the question we which posed in the 
section title.

\section{Is the Confinement property of Yang-Mills theory understandable in terms of classical configurations?}
A few years ago Campostrini et al \cite{digiacomo} observed that the string tension persisted after a few cooling steps,
 and the errors reduced considerably.
This could be interpreted as  evidence in favour of an affirmative answer to the section title question, since the smooth cooled 
configurations account for the value of the string tension. However, there are two facts that  eliminate this conclusion. First, the value of the effective string tension as measured 
from  different observables was seen to decrease when one continues to cool the configurations. Furthermore, Teper \cite{teper}
pointed out that the persistence of the string tension under a few coolings is a result of the local nature of the cooling algorithm.
 However, cooling is known to affect classical configurations, producing for example the 
annihilation of opposite topological charge objects. 
In a previous work \cite{gauge}, we observed  that there is a very strong correlation between the value of  the effective
string tension and the  density of action density peaks $D_{peaks}$, in such a way that the quantity
 $K = \sigma / \sqrt{D_{peaks}}$ stays constant with cooling. This is precisely what we would expect from a classical model 
like ours. Our previous work was based on thermalised data with Wilson action and  $\beta=2.325$ obtained on an $8^3 \times 64$ lattice with twisted boundary conditions. In what follows we will report our new results in this respect.

   We generated  199 SU(2) Yang-Mills configurations on a $16^3 \times 8$ lattice, and periodic boundary conditions, thermalised
 using the Wilson action and  $\beta=2.325$. The configurations were separated by 500 Monte Carlo sweeps after 5000 
initial sweeps.  We have measured Wilson loops and Creutz ratios on them.
Our main results are the following:

1. The Creutz ratios obtained decrease strongly with cooling for all three representations. For example,
for the fundamental representation the Creutz ratio for R=T=6 equals 0.107(3), 0.078(1), 0.045(1), 0.0282(4)
for 10, 20, 50 and 100 coolings. However,
as predicted by our interpretation of this drop, if we compute the quantity $K(6)=\chi(6,6) / \sqrt{D_{peaks}}$ we get   2.22, 2.32, 2.00 and 1.70. Notice that
a drop by a factor 3.8 in $\chi(6,6)$ only produces a $24\% $ difference in K(6).  The actual value is also similar to the one obtained 
for twisted boundary conditions. 

2. One of the most salient  conclusions of our data is that the ratio of Creutz ratios of   different representations does not
depend on cooling. In Fig.~2 we show our results  for $r_j(5) \equiv \chi_j(5,5)/\chi_{\frac{1}{2}}(5,5)$.  The value is a bit smaller than
 the prediction ( See Ref.\cite{greensite}) of 8/3 and 5. Using other values of R=T one   also gets very good constancy
with cooling and values which decrease with increasing R. For example at 80 cooling steps we have $r_1(R)$ equals 
2.33(5), 2.25(6), 2.19(9) and 2.11(36) for R= 4, 5, 6, and 7 respectively. For $r_{\frac{3}{2}}(R)$ the values are 3.94(9), 3.72(13), 3.53(29) for R=4, 5 and 6.
On the basis of our results we are tempted to propose  that these ratios coincide with the uncooled ones.

In conclusion, we have verified our previous findings concerning the correlation  between the string tension and the  
action density local maxima, which reinforces the evidence for Confinement being related to the underlying classical configurations. 
It is important to see that the cooled classical configurations preserve the proportion of Creutz ratios among the different 
representations.

\begin{figure}
\begin{center}
\leavevmode
\epsfverbosetrue
\epsfxsize=200pt
\epsfysize=100pt
\epsffile{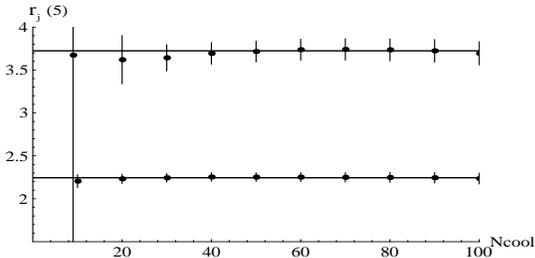}
\caption{The quotient  of Creutz ratios  for R=T=5 ($r_{\frac{3}{2}}(5) \equiv \chi_{\frac{3}{2}}/\chi_{\frac{1}{2}}$ (top) and 
 $r_{1}(5) \equiv \chi_{1}/\chi_{\frac{1}{2}}$ (bottom)) as a function of the number of cooling steps. The horizontal lines are meant to 
guide the eye.}
\end{center}
\end{figure}

\section{Other questions}

The final big question would be: what is the correct description of the  confining classical configurations? 
We cannot at the present moment distinguish between $Q=1/2$ and  $Q=1$ peaks at an individual basis. We can 
repeat our analysis of Ref. \cite{gauge} and compute for each configuration the ratio of the total action divided by 
$8 \pi^2 N_{peaks}$. This should be 1 if all peaks were instantons.  What we found is that the ratio is 0.46, 0.54, 0.61 and 0.65
for 10, 20, 50 and 100 cooling steps. For the unthermalised configurations our values are: 0.66, 0.71, 0.71 and 0.69.


\begin{thebibliography}{9}



\bibitem{invest}
M. Garc\'{\i}a P\'erez, A. Gonz\'alez-Arroyo and P. Mart\'{\i}nez, {\em Nucl. Phys. B (Proc. Suppl.)}  34 (1994), 228.; 
A. Gonz\'alez-Arroyo and P. Mart\'{\i}nez, {\em Nucl. Phys. }  B459 (1996), 337.

\bibitem{greensite} J. Greensite, hep-lat/9607053, Plenary talk at Lat-96. These Proceedings.


\bibitem{doclass}
A. Gonz\'alez-Arroyo and A. Montero, hep-th/9604017.

\bibitem{overimprove}
M. Garc\'{\i}a P\'erez, A. Gonz\'alez-Arroyo, J. Snippe and P. van Baal, {\em Nucl. Phys. }  B413 (1994), 535.

\bibitem{gauge}
A. Gonz\'alez-Arroyo, P. Mart\'{\i}nez and A. Montero,  {\em Phys. Lett. } B359 (1995), 159.

\bibitem{digiacomo}
M. Campostrini et al, {\em Phys. Lett. } 225B (1989), 403.

\bibitem{teper}
M. Teper, {\em Nucl. Phys. } B411 (1994), 855.



\end{thebibliography}
\end{document}